\documentstyle[12pt,epsf]{article}
\begin{document}

\def\eg{{\it e.g.}}\def\etc{{\it etc.}}
\def\etal{{\it etal.}}

\def\lsim{{\buildrel < \over\sim}}
\def\gsim{{\buildrel > \over\sim}}
\def\MSbar{\overline{\rm MS}}
\def\bea{\begin{eqnarray}}
\def\be{\begin{equation}}
\def\bear{\begin{eqnarray*}}
\def\eea{\end{eqnarray}}
\def\ee{\end{equation}}
\def\eear{\end{eqnarray*}}
\def\to{\rightarrow}
\def\ie{{\it i.e.}}
\def\anti{\overline}
\def\pbi{~{\rm pb}^{-1}}
\def\fbi{~{\rm fb}^{-1}}
\def\fb{~{\rm fb}}
\def\pb{~{\rm pb}}
\def\ev{\,{\rm eV}}
\def\mev{\,{\rm MeV}}
\def\gev{\,{\rm GeV}}
\def\tev{\,{\rm TeV}}
\def\wh{\widehat}
\def\wt{\widetilde}
\def\lra{\leftrightarrow}
\def\ra{\rightarrow}
\def\mhalf{m_{1/2}}
\def\gl{\wt g}
\def\mt{m_t}
\def\mb{m_b}
\def\mz{m_Z}
\def\mw{m_W}
\def\mgut{M_U}
\def\mstring{M_S}
\def\wp{W^+}
\def\wm{W^-}
\def\wpm{W^{\pm}}
\def\wmp{W^{\mp}}
\def\q{$q$}
\def\qbar{$\bar{q}$}
\def\g{$g$}
\def\dc{$\delta_c$}
\def\as{\alpha_s}
\def\AmS{{\protect\the\textfont2
  A\kern-.1667em\lower.5ex\hbox{M}\kern-.125emS}}
\def\mathswitchr#1{\relax\ifmmode{\mathrm{#1}}\else$\mathrm{#1}$\fi}
\newcommand{\PB}{\mathswitchr B}
\newcommand{\PW}{\mathswitchr W}
\newcommand{\PZ}{\mathswitchr Z}
\newcommand{\Pg}{\mathswitchr g}
\newcommand{\PH}{\mathswitchr H}
\newcommand{\Pe}{\mathswitchr e}
\newcommand{\Pne}{\mathswitch \nu_{\mathrm{e}}}
\newcommand{\Pane}{\mathswitch \bar\nu_{\mathrm{e}}}
\newcommand{\Pnmu}{\mathswitch \nu_\mu}
\newcommand{\Pd}{\mathswitchr d}
\newcommand{\Pf}{f}
\newcommand{\Ph}{\mathswitchr h}
\newcommand{\Pl}{l}
\newcommand{\Pu}{\mathswitchr u}
\newcommand{\Ps}{\mathswitchr s}
\newcommand{\Pb}{\mathswitchr b}
\newcommand{\Pc}{\mathswitchr c}
\newcommand{\Pt}{\mathswitchr t}
\newcommand{\Pp}{\mathswitchr p}
\newcommand{\Pq}{\mathswitchr q}
\newcommand{\Pep}{\mathswitchr {e^+}}
\newcommand{\Pem}{\mathswitchr {e^-}}
\newcommand{\Pepm}{\mathswitchr {e^\pm}}
\newcommand{\Pmum}{\mathswitchr {\mu^-}}
\newcommand{\PWp}{\mathswitchr {W^+}}
\newcommand{\PWm}{\mathswitchr {W^-}}
\newcommand{\PWpm}{\mathswitchr {W^\pm}}

\def\mathswitch#1{\relax\ifmmode#1\else$#1$\fi}
\newcommand{\MB}{\mathswitch {M_\PB}}
\newcommand{\Mf}{\mathswitch {m_\Pf}}
\newcommand{\Ml}{\mathswitch {m_\Pl}}
\newcommand{\Mq}{\mathswitch {m_\Pq}}
\newcommand{\MV}{\mathswitch {M_\PV}}
\newcommand{\MW}{\mathswitch {M_W}}
\newcommand{\hMW}{\mathswitch {\hat M_\PW}}
\newcommand{\MWpm}{\mathswitch {M_\PWpm}}
\newcommand{\MWO}{\mathswitch {M_\PWO}}
\newcommand{\MA}{\mathswitch {\lambda}}
\newcommand{\MZ}{\mathswitch {M_Z}}
\newcommand{\MH}{\mathswitch {M_\PH}}
\newcommand{\Me}{\mathswitch {m_\Pe}}
\newcommand{\Mmy}{\mathswitch {m_\mu}}
\newcommand{\Mpi}{\mathswitch {m_\pi}}
\newcommand{\Mta}{\mathswitch {m_\tau}}
\newcommand{\Md}{\mathswitch {m_\Pd}}
\newcommand{\Mu}{\mathswitch {m_\Pu}}
\newcommand{\Ms}{\mathswitch {m_\Ps}}
\newcommand{\Mc}{\mathswitch {m_\Pc}}
\newcommand{\Mb}{\mathswitch {m_\Pb}}
\newcommand{\Mt}{\mathswitch {m_\Pt}}
\catcode`@=11
\def\Biggg#1{\hbox{$\left#1\vbox to 22.5\p@{}\right.\n@space$}}
\catcode`@=12
\newcommand\refq[1]{$^{#1}$}
\newcommand\ind[1]{_{\rm #1}}
\newcommand\aopi{\frac{\as}{\pi}}
\newcommand\oot{\frac{1}{2}}
\newcommand\sinsthw{\sin^2(\theta_{\rm W})}
\newcommand\logmu{\log(\mu^2/\Lambda^2)}
\newcommand\Lfb{\Lambda^{(5)}}
\newcommand\Lfc{\Lambda^{(4)}}
\newcommand\Lf{\Lambda_5}
\newcommand\epem{\ifmmode e^+e^- \else $e^+e^-$ \fi}
\newcommand\mupmum{ \mu^+\mu^- }
\newcommand\msbar{\ifmmode{\overline{\rm MS}}\else $\overline{\rm MS}$\ \fi}
\newcommand\Q[1]{_{\rm #1}}
\newcommand\pplus[1]{\left[\frac{1}{#1}\right]_+}
\newcommand\plog[1]{\left[\frac{\log(#1)}{#1}\right]_+}
\newcommand\sh{\hat{s}}
\newcommand\epbar{\overline\epsilon}
\newcommand\nf{\alwaysmath{{n_{\rm f}}}}

\newcommand{\aem}{\alpha_{\rm em}}
\newcommand{\nlf}{\alwaysmath{{n_{\rm lf}}}}
\newcommand{\ep}{\epsilon}
\newcommand{\aop}{\frac{\as}{2 \pi}}
\newcommand{\Tf}{{T_{\rm f}}}
\newcommand{\mub}{\ifmmode \mu{\rm b} \else $\mu{\rm b}$ \fi}
\newcommand\alwaysmath[1]{\ifmmode #1 \else $#1$ \fi}
\newcommand{\TeV}{{\rm TeV}}
\newcommand{\GeV}{{\rm GeV}}
\newcommand{\MeV}{{\rm MeV}}
\newcommand{\LQCD}{\ifmmode \Lambda_{\rm QCD} \else $\Lambda_{\rm QCD}$ \fi}
\newcommand{\LMSB}{\ifmmode \Lambda_{\overline{\rm MS}} \else
          $\Lambda_{\overline{\rm MS}}$ \fi}
\newcommand{\qb}{\overline{q}}
\def\pp{\ifmmode p\bar{p} \else $p\bar{p}$ \fi}
\def\VEV#1{\left\langle #1\right\rangle}
\def\LMSb{\ifmmode \Lambda_{\rm \overline{MS}} \else
$\Lambda_{\rm \overline{MS}}$ \fi}
\def\ie{\hbox{\it i.e.}{}}      \def\etc{\hbox{\it etc.}{}}
\def\eg{\hbox{\it e.g.}{}}      \def\cf{\hbox{\it cf.}{}}
\def\etal{\hbox{\it et al.}}
\def\dash{\hbox{---}}
\def\abs#1{\left| #1\right|}   

 
\begin{titlepage}
\begin{flushright}
ETH-TH/2000-05\\
March 1999
\end{flushright}
\begin{center}
\vspace*{2.5cm}
{\Large\bf
Electroweak Physics, Theoretical Aspects
} 
\vskip 1cm
{\large Z. Kunszt} \\
\vskip 0.2cm
{\it 
Institute for Theoretical Physics, ETHZ\\
 CH-8093 Zurich, Switzerland
} \

\vskip 4cm
\end{center}
 
\begin{abstract}
\noindent
I discuss two aspects of the electroweak interactions:
the status of the precision measurement of the electroweak
parameters and their impact on the Higgs search at future
colliders.

\end{abstract}

\vfill
\noindent\hrule width 3.6in\hfil\break
{ Plenary Talk at the UK Phenomenology
Workshop on Collider Physics, Durham, 1999}
\hfil\break

\end{titlepage}
\setcounter{footnote}{0}
 
\newpage

\section{Precision calculations}

During the last decade we witnessed an impressive progress
 at LEP, SLC and Tevatron achieved by collecting an enormous amount of 
electroweak data on the  Z and W  bosons  and their interactions \cite{Swartz:1999xv,Sirlin:1999zc}.
This allows  for  an unprecedented precision test
 of the Standard Model at the level of the per mil accuracy.
At this precision  one
 and two-loop
quantum fluctuations give measurable contributions and  an interesting 
upper limit on 
the mass of the Higgs-boson can be obtained.

\subsection{Input values}

In the Standard Model at tree level  the gauge bosons
$\gamma,W,Z$ and  their interactions 
are described in terms of three parameters:
 the two gauge coupling constants  $g,g^{\prime}$ and
 the vacuum expectation value of the Higgs-field $v$.  We need to know their values
as precisely as possible.
They have to be fitted to  the three  best measured physical quantities
of    smallest 
experimental error: 
 $G_{\mu},\MZ$ and $\alpha$. The muon coupling 
$G_{\mu}$ is extracted from the precise measurement of the muon life-time
using   the theoretical expression 
\begin{equation}\label{Gmu}
\frac{1}{\tau_{\mu}}=\frac{G^2_{\mu}m_{\mu}^5}{192\pi^3}\left(1-\frac{8m_e^2}
{m^2_{\mu}}\right)\\ \nonumber
\left[1+1.810\left(\frac{\alpha}{\pi}\right) + 
(6.701\pm 0.002)\left(\frac{\alpha}{\pi}\right)^2+ ...\right]
\end{equation}
It is crucial
 that   that the  electromagnetic
corrections  to the muon life time are finite and that 
they are known up to next-to-leading order accuracy.
The ${\cal O}(\aopi)$ term
has been obtained   by van Ritbergen and Stuart only very recently \cite{vanRitbergen:1999fi},
it lets  reduce   the theoretical
error with factor of two. 
Equation \ref{Gmu}  gives  a unique correspondence between the muon
life-time
 and $G_{\mu}$ since the non-photonic corrections are 
all  lumped into its definition. As a result  
$G_{\mu}$ can be considered as a  physical quantity.
Using the measured value \cite{PDGcaso} we get
\begin{equation}
G_{\mu}=(1.16637\pm 0.00001)\times 10^{-5}\gev^{-2}\,.
\end{equation}
The  value of $\MZ$
 is extracted  from the line shape measurement at the $Z$-pole.
There are subtleties in the theoretical definition of the mass
and the width at higher order associated with the truncation
of the perturbative series and gauge invariance. The latest best value
is \cite{EWWG2000}
\begin{equation}
\MZ=(91.1871\pm 0.0021)\gev\,.
\end{equation}
Finally, the best value of $\alpha$ is extracted 
 from the precise measurement of the electron anomalous
magnetic moment $(g_{\rm e}-2)$~\cite{PDGcaso}
\begin{equation}
{1}/{\alpha }=137.03599959\pm 0.00000038\,.
\end{equation}
I recall the  leading order relations 
\begin{equation}
G_{\mu}=\frac{1}{\sqrt{2}v}\,,\quad 
\MZ=\frac{1}{2\cos\theta_W}gv\,,\quad 
\alpha=\frac{g^2}{4\pi}\sin^2\theta_W\,,\quad  
\tan\theta_W=\frac{g^{\prime 2}}{g^2}\,.
\end{equation}
Additional  physical quantities
like the mass of the W-boson $\MW$, the lepton asymmetries at the
Z-pole, the leptonic width of the Z-boson $\Gamma_{\rm l}$ \etc
are derived quantities.
At the level of the per mil accuracy the predictions
obtained in Born approximations for derived
quantities, however, fail significantly.

\subsection{Quantum corrections}

The precision test of the Standard Model
is obtained by confronting the measured values of
 derived quantities with the precise
prediction of the theory.
Since the Standard Model
is a renormalizable quantum field theory \cite{tHV} 
the theoretical 
 predictions of the theory can be improved systematically 
by calculating   higher order corrections.
In particular, the recent  precision of the data
 requires  the study  
of the complete  next-to-leading
order corrections,  resummation of  large logarithmic
contributions  and  a number of 
 two loop corrections. 
 At higher order 
the derived quantities  show sensitivity also to  
the values of the mass parameters  $\mt,\MH, \mb$ and the QCD
coupling constant
$\alpha_s$.
From direct measurements one obtains 
$\alpha_s=0.119\pm0.002$,
$\mt=173.8\pm5.0\gev$ and $m_b=4.7\pm 0.2\gev$, $\MH \ge 102\GeV$.
The error bars give    parametric
uncertainties in the predictions and  limit our
ability to extract a   precise value of the Higgs mass.
The calculation of the higher orders requires 
 a choice of the renormalization scheme.
The on-shell scheme 
can be regarded as the  extension of the well-known  scheme
of renormalization in QED, it 
uses as input $\alpha$, 
$\MZ$, $\MW$, 
$\MH$ and $m_f$. 
 In the
 $\overline{\rm MS}$-scheme the measured values of
$\alpha$, $G_{\mu}$, $\MZ$, $m_f$, $\as$
 are used to fix the input parameters of the theory with  $\MH$  as free parameter.
The $\MSbar$ gauge couplings evaluated 
at the scale of $\MZ$ 
are denoted as $\hat{e}$ and $
\hat{s}^2 =\sin^2\hat{\theta}_W(\MZ)$.
 The on-shell
definition of the mixing angle  
$s^2=\sin^2\theta$
 is given by  the tree level relation
$s^2=1-\MW^2/\MZ^2$ and, therefore, it is a physical quantity. 
The renormalized parameters $\hat{s
}^2$, $\hat{e}^2$ can be completely
calculated  in terms of $G_{\mu}$, $\alpha$ and $\MZ$.
It is customary to define auxiliary dimensionless parameters.
 $r_W$ is defined by the relation 
\be
s^2c^2\equiv \frac{\pi\alpha}{\sqrt{2}G_{\mu}\MZ^2(1-\Delta r_W)}\,.
\ee
It is  a physical quantity and gives the radiative corrections
to the $\MW$.
The asymmetries measured at the Z-pole are 
given in term of the effective mixing angle
\be
\sin^2\theta^{eff}_W=\frac{1}{4}\left(1-\frac{\bar{g}_{Vl}}{\bar{g}_{Al}}\right)
=s^2(1+\Delta k^{'}) 
\,,\quad 
A_{\rm FB}^l\equiv \frac{3\bar{g}^{2}_{Vl}\bar{g}^{2}_{Al}}{
(\bar{g}^{2}_{Vl}+\bar{g}^{2}_{Al})^2}
\ee
where the dimensionless parameter $\delta k^{'}$ is again
defined in terms of physical quantities.
The leptonic width depends on the vector axial vector
coupling and on the  corrections to the
$Z$-propagator. This requires the introduction of the so called
 $\rho$-parameter
\be
\Gamma_l=
\frac{G_{\mu}\MZ^3 }
{6\pi\sqrt{2}}\left(\bar{g}_{Vl}^2+\bar{g}_{Al}^2\right)
\rho
\ee
These type of auxiliary functions can be calculated 
in  different renormalization schemes.
The corrections $\Delta r_W$,  $\Delta k'$, $\Delta \rho$  
    (and a number of
additional useful dimensionless quantities) are known in various 
schemes and play an important role in the analysis of
electroweak physics, because they give  the 
precise predictions of the theory for simple observables as 
$\MW$, the leptonic
asymmetries \etc in terms of $\alpha,G_{\mu}$ and $\MZ$.
It is very useful to have the results 
in different schemes since it allows for cross-checking
the correctness of the result  and to  estimate
 the remaining theoretical errors given by the missing higher order
contributions.  
The electroweak radiative corrections are dominated by two leading
contributions: the running of the electromagnetic
coupling and large  
$\mt$ effects to $\rho$
$( \Delta \rho_t\approx 3G_{\mu}\mt^2/(8\pi^2\sqrt{2}))$.

\subsection{Running electromagnetic coupling}

Because of  gauge invariance the running of $\alpha$ is completely
given by the photon self-energy contributions
\begin{equation}
\alpha(\MZ)=\frac{\alpha}{1-\Delta\alpha}
\end{equation}
where
\begin{equation}
\Delta\alpha=
-{\rm Re}\left({\hat\Pi}^{\gamma}(\MZ^2)\right)=
-{\rm Re}\left({\Pi}^{\gamma}(\MZ^2)\right)+
{\rm Re}\left({\Pi}^{\gamma}(0)\right)\,.
\end{equation}
The self-energy contribution is large
($\approx 6\%$). It can be split into leptonic
and hadronic contributions
\begin{equation}
\Delta\alpha=\Delta\alpha_{\rm lept}
+\Delta\alpha_{\rm had}
\end{equation}
The leptonic part is known up to three loop 
\begin{equation}
\Delta\alpha_{\rm lept}=314.97687(16)\times 10^{-4}
\end{equation}
and the remaining theoretical error is completely negligible.
The hadronic contribution is more problematic since it  
can not be calculated theoretically with the required precision
since the light quark loop contributions have non-perturbative
QCD effects. One can extract it, however, from the data using the
relation
\bea
\Delta\alpha_{\rm had}&=&\frac{\alpha}{3\pi}\MZ^2{\rm Re}
\int_{4m^2_{\pi}}^{\infty}\frac{R_{\epem}(s^{\prime)}}{s^{\prime}(
s^{\prime}-\MZ^2-i\epsilon)}\nonumber\\
R_{\epem}(s)&=&\frac{
\sigma(
\epem\to\ \gamma^*\to \ {\rm hadrons})}
{\sigma(
\epem\to\ \mupmum)}
\eea
Conservatively,  one calculates the high energy $\sqrt{s}\ge 40\gev$
contribution using perturbative QCD and  the low energy contribution
 $\sqrt{s}\le 40\gev$
is estimated using data \cite{Jegerlehner:1999hg}.
 Unfortunately, the precision 
of the low energy data is not good enough and the error  from this
source  dominates the error of the theoretical predictions
\be
\Delta^{(5)}\alpha_{\rm had}=0.02804\pm0.00064\,,\\
\alpha^{-1}(\MZ)=128.89\pm0.09\,.
\ee
One can, however,  achieve a factor of three
reduction of the estimated error
assuming that the theory can be used 
down to $\sqrt{s}=m_{\tau}$ when
quark mass effects can be included up to three loops.
  Such an analysis
is quite well motivated  by  the successful
results on the   tau life-time. In the hadronic vacuum polarization
 the non-perturbative power corrections appear to be  suppressed
and the unknown  higher order perturbative
contributions are relatively small. In this theory
driven approach the error is reduced 
to an acceptable $0.25\% $ value
\be
\alpha^{-1}(\MZ)=128.905\pm0.036\,.
\ee
It is unlikely that the low energy hadronic total 
cross section will be measured in the
foreseeable future 
with a precision leading to essential improvement.

\subsection{Comment on the muon  anomalous magnetic moment}

I note that  the hadronic vacuum polarization contribution
to the muon anomalous magnetic moment
$a_{\mu}\equiv (g_{\mu}-2)/2$
 is more problematic as a result of 
 a different weight factor in the dispersion integral
\bea
a_{\mu}^{\rm had}&=&
\left(\frac{\alpha \Mmy}{3\pi}\right)
\Biggl[
\int^{E^2_{\rm cut}}_{4\Mpi^2}\frac{R_{\epem}(s)^{\rm data}(s)K(s)}{s^2}
\nonumber \\
&&+\int^{\infty}_{E^2_{\rm cut}}\frac{R_{\epem}(s)^{\rm pQCD}(s)K(s)}{s^2}
\Biggr]
\eea
where we splitted the perturbative and low energy contributions. $K(s)$
 is a kinematical weight factor which together with $1/s^2$
 enhances
the low energy contributions. As a result
the experimental error of the measured
value of the  hadronic vacuum polarization contribution
leads   more than  $1\%$ error in the theoretical prediction.
It is expected that high statistics 
data collected in ${\rm DA}\Phi{\rm NE}$ in the future will 
reduce this error with a factor of two. Such an improvement is
very well motivated in view of the experimental 
effort of  the ongoing Brookhaven
experiment which will  achieve a precision of
 $\approx 40 \times 10^{-11}$, a significant reduction
in comparison with the present error of $\approx 730 \times 10^{-11}$ 
(see \cite{Czarnecki:1998nd} and references therein).
Note, however, that the hadronic contribution from light-to-light scattering
  diagrams cannot be measured and the  theoretical estimates have
large uncertainties leading to a 
theoretical error of $\approx 40 \times 10^{-11}$. 
Accepting this estimate with the precise measurement of 
$a_{\mu}$ it will  be possible to test for the present of
anomalous coupling (SUSY) contributions provided they are large
( $\approx 100 \times 10^{-11}$ or larger).
It is unlikely that one gets improvements over the existing
LEP limits.

\begin{figure}[htbp]
   \vspace{-3.8cm}
   \epsfysize=16cm
   \epsfxsize=16cm
   \centerline{\epsffile{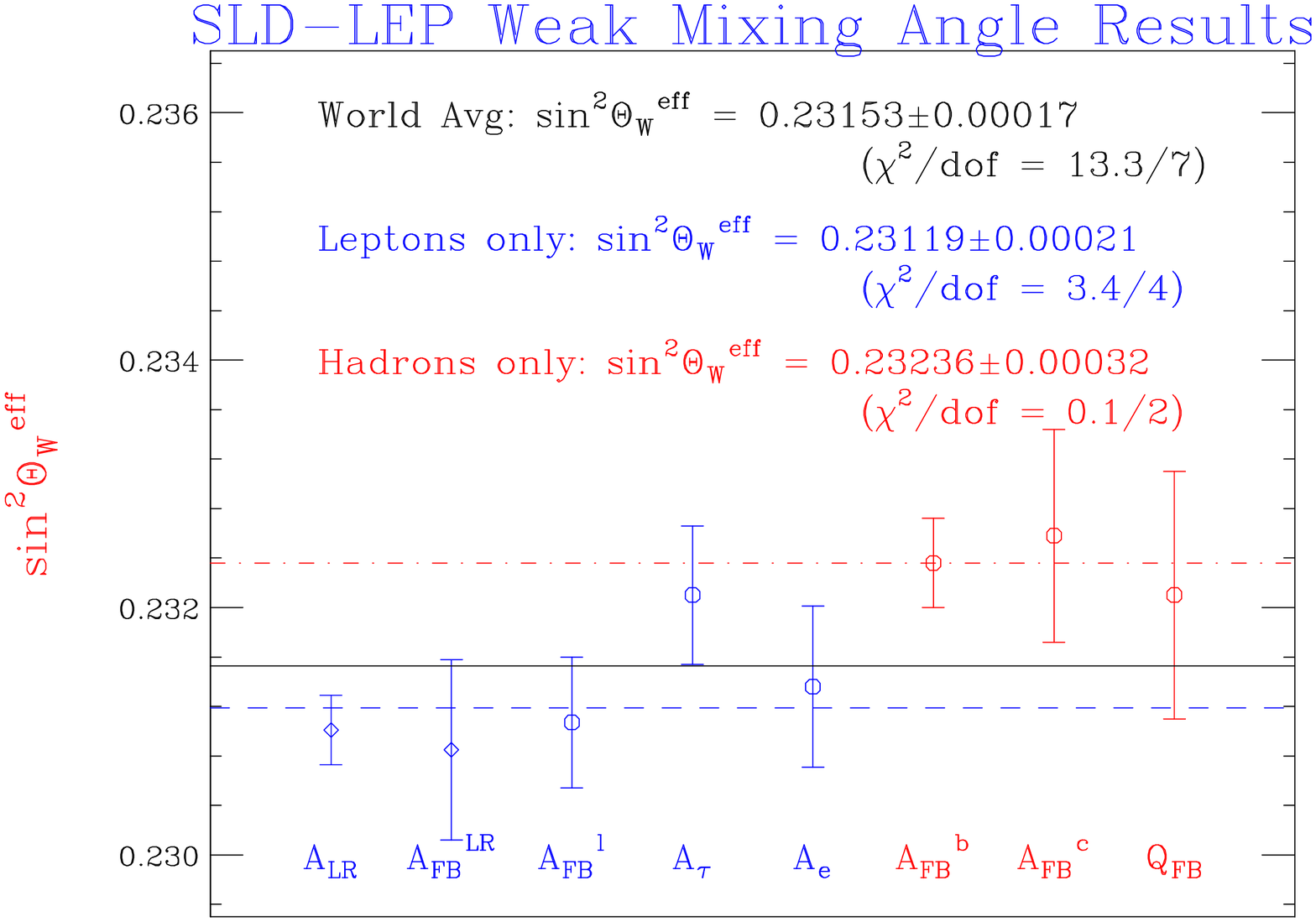}}
   \vspace*{-2.0cm}
\caption[dummy] { \ \ 
Summary of all the determinations of 
  $\sin^2\theta_W^{\rm eff}$ \cite{Swartz:1999xv}.
   \label{fig:diag1} }
\end{figure}

\subsection{Higher order corrections to $\MW$ and the mixing angle }

As we noted above, the simplest physical observables for  
precise test of the Standard Model are
$\MW$ and the $\sin^2\theta^{eff}_W$. 
It is convenient to consider the radiative corrections
in the $\MSbar$ scheme where with good accuracy
$\sin^2\theta^{eff}_W\approx \hat{s}^2$.
It is given in terms of the input parameters
via the relation
\be
\hat{s}^2\hat{c}^2=\frac{\pi\alpha(\MZ)}{\sqrt{2}G_{\mu}\MZ(1-\hat{r}_w)}
\ee
where  $\hat{r}_w=0$ in leading order. Using the measured value of
$\sin^2\theta^{eff}_W,\MZ$ and $G_{\mu}$ we obtain a value
$\hat{r}_{W}=0.0058\pm 0.000480$  different from zero at the
$12\sigma$
level. If one carries out a similar analysis for $\MW$ the evidence for
the presence of subleading corrections  is even better.
 The radiative correction 
$\hat{r}_W$ does not contain the large
effect from the running $\alpha$ but it receives  large
custodial symmetry  violating corrections because of  the large
top-bottom mass splitting
\be
\Delta \hat{r}_W|_{\rm top}
=-c^2/s^2\Delta \rho \approx  
0.0096  \pm 0.00095
\ee
Subtracting this value we get   about $6\sigma$ difference coming
from the  the loops involving the bosonic
sector (W,Z,H) and subleading fermionic contributions. 
At this level of
accuracy many other corrections start to become important and  the
the size of  errors coming from the errors in the input parameters
leads to effects of the same order. In particular, we get some
sensitivity to 
the value of the Higgs-mass. 
 Beyond the complete one
loop corrections it was possible to evaluate    
all  important two loop corrections: 
  ${\cal O}(\alpha^2\ln(\MZ/\Mf)$ corrections 
 with light fermions, mixed electroweak QCD corrections
of ${\cal O}(\alpha\alpha_s)$,
 two loop electroweak corrections
 enhanced by top mass effects of
  ${\cal O}(\alpha^2(\mt^2/\MW^2)^2)$ together with 
 the subleading parts of  
 ${\cal O}(\alpha\alpha_s^2 \mt^2/\MW^2)$ and
the very difficult subleading correction
 of  ${\cal O}(\alpha^2 \mt^2/\MW^2)$.
 It is remarkable that last  contribution proved to be
 important in several respect \cite{Degrassi:1999jd}.
Its inclusion reduced significantly
 the scheme dependence of the results and lead to a  significant  
 reduction of the upper limit on the Higgs mass.

\subsection{Global fits}
 This summer the LEP experiments and SLD could finalize their results
 on the electroweak precision data. The most important development
 is that the final value of SLD on the leptonic polarization asymmetry was reported
 which implies $\sin^2\theta^{eff}_W=0.23119\pm 0.00020$.
 A nice summary of the results is given in Figure 1 \cite{Swartz:1999xv}\,.
According to a recent analysis of the EWWW working group \cite{EWWG2000}
the new world average is
\be
\sin^2\theta^{eff}_W=0.23151\pm 0.00017 {\rm \ \ \ with\ \ \ }
 \chi^2/{\rm d.o.f=13.3/7} \,.
\ee
 This gives only rather low  confidence level of $6.4\%$.
 The origin of this unsatisfactory result is the $2.9\sigma$ discrepancy
 between the values  $\sin^2\theta^{eff}_W$ derived from the SLAC 
 leptonic polarization asymmetry data and from the forward backward asymmetry
 in the b-b channel at LEP and SLC.
 The results obtained from a
 global fit to all data give somewhat better result but there we are hampered
 with the problem that the polarization asymmetry parameters
 disagree with each other with $2.7\sigma$, therefore, the $\chi^2$ is
 relatively large.

\subsection{The weak charge of the atomic Cesium}
Recently, a new determination of the weak charge of the atomic
Cesium (via studying the $6s\to 7s$ parity violating tensor transition)
has been presented \cite{Wood:1999xx}
\be
Q_W(^{133}_{\ 55} Cs)=-72.06\pm (0.28)_{exp}\pm(0.34)_{th}
\ee
with considerable improvement with respect to earlier
results
\be
Q_W(^{133}_{\ 55} Cs)=-71.04\pm (1.58)_{exp}\pm(0.88)_{th}\,.
\ee
In the theory, $Q_W$ measures the product
of the vector and axial vector neutral current coupling
of the u and the d quarks $C_{1u},C_{1d}$
\be
Q_W=-2\left[C_{1u}(2Z+N)+C_{1d}(Z+2n)\right]
\ee
A crucial feature of this test is that
it constrains the value of  the parameter $\epsilon_3$ \cite{altbarb}
\be
Q_W=-72.87\pm0.13-102\epsilon_3^{\rm SM}
\ee
with $\epsilon_3^{\rm SM}=0.0053 \pm 0.0013$
for $\MH=70-1000\gev$. According to the data \cite{Casalbuoni:1999yy}
\be
Q_W^{exp}-Q_W^{th}=1.28\pm 0.46
\ee
a three standard deviation effect. 
One should accept this result with some care
in view of the 
 significant reduction of the experimental error.
It would be important to cross-check this result with 
with other independent experiments. 
Also the estimate of the theoretical uncertainties
coming from atomic physics calculation may be too optimistic.
In ref.  \cite{Casalbuoni:1999yy}
 the deviation was attributed
to the existence of a non-sequential $Z^{'}$-boson.
and the data have been used to constraint its properties.

\section{Constraints on the Higgs mass}

\subsection{Upper limit from the measured value of $\MW$}
 The final  results of the electroweak radiation
 corrections for $\MW$ and $\sin^2\theta^{eff}_W$
 can be parameterized 
  in terms of the input parameters
 including their errors in simple approximate analytic form 
\cite{Degrassi:1999jd}.
 For example in the $\overline{MS}$-scheme one obtains for the W-mass
\bea\label{MWcorrfit}
\MW&=&80.3827-0.0579\ln (\frac{\MH}{100})
           -0.008\ln^2 (\frac{\MH}{100})\nonumber\\
          && -0.517\left(\frac{\delta\alpha_h^{(5)}}{0.0280}-1\right)
           +0.543\left[\left(\frac{\mt^2}{175}-1\right)\right]\nonumber\\
          && -0.085 \left(\frac{\alpha_s(\MZ)}{0.118}-1\right)
\eea
where $\mt$, $\MH$ and $\MW$  are in $\gev$ units.
 This formula accurately reproduce the result obtained with 
 numerical evaluation of all corrections in the range 
 $75\gev \le \MH \le 350 \gev$ with  maximum deviation of less than
$1\mev$. Using  the world average of the
measured values  of the W-boson mass $\MW=80.394\pm  0.042\gev$ \cite{monig}
(with input parameters $\alpha_s=0.119\pm 0.003$, $\mt=174.3\pm 5.1\gev$,
$\delta\alpha^{(5)}=0.02804\pm 0.00065$
 one  obtains at $95\%$ confidence level an allowed range for
the Higgs mass of $73\gev \le \MH\le 294\gev$.
 Similar results exists also for $\sin^2\theta^{ eff}_W$ extracted from
 the asymmetry measurements at the Z-pole with somewhat better (95\% confidence
 ) limits
 of $95\gev \le \MH\le 260\gev$.
 Without global fits we got a semi-analytic insight on  the sensitivity 
 of the precision tests to the
Higgs mass. We also see that the precise measurements 
of $\MW$
have already provided us with
 competitive values  in comparison with the those obtained
from the measurement of  $\sin^2\theta^{ eff}_W$.

\subsection{Results from global fits}
It is interesting that the values of the Higgs mass obtained
in a recent global fit \cite{D'Agostini:2000ws}
 are in good agreement  with the simple analysis based on the value of
of $\MW$ or $\sin^2\theta^{eff}_W$  as described above.
From the global fit one obtains an expected value
for the Higgs boson of $160-170\gev$ with error of $ \pm 50-60 \gev$.
The 95\% confidence level upper limit is about
$260-290\gev$.

\subsection{Can the Higgs-boson be heavy?}

The precision data can not rule out yet 
 dynamical symmetry breaking with some heavy Higgs-like
scalar and vector resonances.
The minimal model to describe this alternative  is obtained
by assuming that the new particles are heavy (more than 0.5 \tev)
 and the linear $\sigma$-model Higgs-sector
of the Standard Model is replaced 
 by the non-renormalizable
non-linear $\sigma$-model. It  can  be derived  also as
 an effective chiral vector-boson Lagrangian with non-linear 
realization of the gauge-symmetry \cite{ApBe80, longhitano:81}.
How can we reconcile this more phenomenological
approach  with the precision
data?
Removing the
Higgs boson from the Standard Model while keeping
the gauge invariance is a  relatively mild
change.  Although the model
becomes non-renormalizable, but at the one-loop
level  the radiative effects grow only logarithmically with the cut-off
at which new interactions should appear.
In equation (\ref{MWcorrfit}) the Higgs-mass is replaced by
this  cut-off 
 The logarithmic terms
are universal, therefore, their coefficients
must remain the same. The  constant terms, however, can be different
from those of the Standard Model. The one loop
corrections of the effective  theory
require the  introduction of  new free parameters which
influence the value of the constant terms. 
The data, unfortunately, do not have sufficient precision
to significantly  constrain  
the constant term appearing in $M_W$, $\sin^2\theta^{eff}_W$
and $\Gamma_l$ (or alternatively in the parameters 
 $\epsilon_1,\epsilon_2,\epsilon_3$~\cite{altbarb} or $S,T,U$~\cite{pestak} ).
In a recent analysis ~\cite{Bagger:1999te} 
it has been
 found that due to the screening
of the symmetry breaking sector \cite{veltman},
 alternative theories with dynamical 
symmetry breaking and heavy scalar and vector bosons 
still can be in 
 agreement  with  the precision data 
up to a cut-off scale of $3\TeV$.



\begin{thebibliography}{999}
\parskip 0pt
\itemsep=0pt

\def\np#1#2#3  {{Nucl. Phys. }{\bf #1}, #2 (19#3)}
\def\nc#1#2#3  {{Nuovo. Cim. }{\bf #1}, #2 (19#3)}
\def\pl#1#2#3  {{Phys. Lett. }{\bf #1}, #2 (19#3)}
\def\pr#1#2#3  {{Phys. Rev. }{\bf #1}, #2 (19#3)}
\def\prl#1#2#3  {{Phys. Rev. Lett. }{\bf #1}, #2 (19#3)}
\def\prep#1#2#3 {{Phys. Rep. }{\bf #1}, #2 (19#3)}
\def\zp#1#2#3  {{Z. Phys. }{\bf #1}, #2 (19#3)}
\def\rmp#1#2#3  {{Rev. Mod. Phys. }{\bf #1}, #2 (19#3)}

\bibitem{Swartz:1999xv}
M.~L.~Swartz,
hep-ex/9912026 and references therein.

\bibitem{Sirlin:1999zc}
A.~Sirlin,
hep-ph/9912227  and references therein.


\bibitem{vanRitbergen:1999fi}
T.~van Ritbergen and R.~G.~Stuart,
hep-ph/9904240; 
Phys.\ Rev.\ Lett.\  {\bf 82} (1999) 488
[hep-ph/9808283].

\bibitem{PDGcaso}
C.~Caso {\it et al.}, Eur.\ Phys.\ J.\  {\bf C3}, 1 (1998).


\bibitem{EWWG2000}
LEP and SLD  Electroweak Working Group, preprint CERN EP/2000-16,

\bibitem{tHV}
G. `t Hooft, \np{B35}{167}{71};
G. `t Hooft and M. Veltman, \np{B44}{189}{72};



\bibitem{Jegerlehner:1999hg}
F.~Jegerlehner,
hep-ph/9901386.


\bibitem{Degrassi:1999jd}
G.~Degrassi and P.~Gambino,
hep-ph/9905472.


\bibitem{Czarnecki:1998nd}
A.~Czarnecki and W.~J.~Marciano,
hep-ph/9810512.



\bibitem{Wood:1999xx}
C.~S.~Wood, S.~C.~Bennett, J.~L.~Roberts, D.~Cho and C.~E.~Wieman,
Can.\ J.\ Phys.\  {\bf 77} (1999) 7.

\bibitem{Casalbuoni:1999yy}
R.~Casalbuoni, S.~De Curtis, D.~Dominici and R.~Gatto,
Phys.\ Lett.\  {\bf B460} (1999) 135
[hep-ph/9905568].

\bibitem{altbarb}
G. Altarelli, R. Barbieri and F. Caravaglios,
\np{B405}{3}{93}

\bibitem{monig}
K. M\"onig, these proceedings

\bibitem{D'Agostini:2000ws}
G.~D'Agostini and G.~Degrassi,
hep-ph/0001269

\bibitem{ApBe80}
 T. Appelquist and C. Bernard Phys.\ Rev.\  {\bf D22}  200  (1980).

\bibitem{longhitano:81}
A. Longhitano, Phys.\ Rev.\  {\bf D22},  1166 (1980), 
\ Nucl.\ Phys. {\bf B188}, 118 (1981).

\bibitem{Bagger:1999te}
J.~A.~Bagger, A.~F.~Falk and M.~Swartz, hep-ph/9908327.
\bibitem{veltman}
M. Veltman,  Act.\ Phys.\ Pol.\ {\bf B8}, 475 (1977).

\bibitem{pestak}
M. Peskin and T. Takeuchi, \prl{65}{2963}{90} .


\end{thebibliography}
\end{document}